\documentclass[a4paper,12pt]{article}
\usepackage{txfonts}
\usepackage[utf8x]{inputenc}

\title{Current Response in Extended Systems as a Geometric Phase:
  Application to Variational Wavefunctions}

\author{Bal\'azs Het\'enyi \\ \\
Department of Physics, Bilkent University \\ 06800, Ankara, Turkey
  \\ and \\ Institute for Theoretical Physics/Computational Physics \\ Graz
  University of Technology \\ Petersgasse 16, Graz, A-8010, Austria}

\begin{document}

\maketitle

\abstract{The linear response theory for current is investigated in a variational
  context.  Expressions are derived for the Drude and superfluid weights for
  general variational wavefunctions.  The expression for the Drude weight
  highlights the difficulty in its calculation since it depends on the exact
  energy eigenvalues which are usually not available in practice.  While the
  Drude weight is not available in a simple form, the linear current response
  is shown to be expressible in terms of a geometric phase, or alternatively
  in terms of the expectation value of the total position shift operator.  The
  contribution of the geometric phase to the current response is then analyzed
  for some commonly used projected variational wavefunctions (Baeriswyl,
  Gutzwiller, and combined).  It is demonstrated that this contribution is
  independent of the projectors themselves and is determined by the
  wavefunctions onto which the projectors are applied. }

\section{Introduction}

Variational studies have contributed greatly to our understanding of
correlated systems.  In part this is due to their relative simplicity,
applicability to larger sizes irrespective of the number of dimensions, and
the easily accessible physical insight they provide.  In the case of the
Hubbard model~\cite{Gutzwiller63,Gutzwiller65,Hubbard63,Kanamori63} frequently
used variational wavefunctions include the Gutzwiller
~\cite{Gutzwiller63,Gutzwiller65} (GWF) and Baeriswyl
wavefunctions~\cite{Baeriswyl86,Baeriswyl00,Dzierzawa97} (BWF), and their
combinations.  The former is based on suppressing charge fluctuations in the
noninteracting solution, the latter on projecting with the hopping operator
onto a wavefunction in the large interaction limit.

The GWF has been studied by a variety of methods.  It can be solved
exactly in one~\cite{Metzner87,Metzner89} and
infinite~\cite{Metzner88,Metzner89,Metzner90} dimensions, and it can be
simulated in two and three dimensions by variational Monte
Carlo~\cite{Yokoyama87}.  The one-dimensional exact solution produces a state
with a finite discontinuity of the momentum density at the Fermi surface.
Millis and Coppersmith~\cite{Millis91} have investigated the response of the
GWF and have concluded that it is metallic with a conductivity proportional to
the kinetic energy.  Insulating behavior in projected wavefunctions similar to
the one due to Gutzwiller can be produced by generalized projection
operators~\cite{Capello05,Capello06}, for example non-centro-symmetric or
singular projectors.

Calculating the Drude or the superfluid weight in a variational context is a
difficult issue.  These two quantities can be cast in terms of identical
expressions (see Eq. (\ref{eqn:Dc})), the second derivative of the ground
state energy with respect to a Peierls
phase~\cite{Kohn64,Shastry90,Scalapino92,Scalapino93}.  As pointed out by
Scalapino, White, and Zhang, the two quantities differ in the interpretation
of the derivative~\cite{Scalapino92,Scalapino93}.  For the Drude weight the
Peierls phase shifts the ground state energy adiabatically, remaining always
in the same state, for the superfluid weight level crossings are also
considered.

In this work general expressions for the Drude and superfluid weights are
derived in a variational setting.  For the Drude weight deriving an easily
applicable expression is a difficult issue, since the expression derived
herein depends on the exact eigenvalues of the perturbed Hamiltonian, in
practical settings often not available.  It is then demonstrated that the
linear response expression for the current can be cast in terms of a geometric
phase.  The tool for calculating this geometric phase (the total position
shift operator) are also presented.  The formalism is then used to interpret
the current response of projected wavefunctions.  It is demonstrated that the
current response in the commonly used Gutzwiller and Baeriswyl projected, as
well as wavefunctions based on combinations of the two projections, produce a
current response identical to the wavefunction on which the projections are
applied (the Fermi sea or the wavefunction in the strongly interacting limit).

\section{Drude and superfluid weights in variational theory}

An expression for the frequency ($\omega$) and wave vector (${\bf
  q}$)-dependent conductivity was derived by Kohn~\cite{Kohn64}.  The DC
conductivity (Drude weight, $D_c$) corresponds to the strength of the
$\delta$-function peak of the conductivity in the zero frequency limit.  The
correct expression for $D_c$ is obtained by first taking the limit (${\bf
  q}\rightarrow 0$) and then the other limit $\omega\rightarrow 0$.  $D_c$ is
often expressed~\cite{Kohn64,Shastry90} in terms of the second derivative of
the ground state energy with respect to a phase associated with the perturbing
field as
\begin{equation}
\label{eqn:Dc}
D_c = \frac{\pi}{L} \left[ \frac{\partial^2 E_0(\Phi)}{\partial
    \Phi^2}   \right]_{\Phi=0}.
\end{equation}
Here $E_0(\Phi)$ denotes the perturbed ground state energy, $\Phi$ denotes the
Peierls phase.

Scalapino, White, and Zhang (SWZ)~\cite{Scalapino92,Scalapino93} have
investigated the distinction between the Drude and superfluid weights.  In
particular they studied the importance of the order of different limits
($\omega\rightarrow 0$, ${\bf q}\rightarrow 0$) for the conductivity.  In a
variational context implementation of the frequency limit is not
straightforward, since, strictly speaking there is no frequency to speak of.
However, SWZ have also pointed out that the derivative with respect to the
phase $\Phi$ in Eq. (\ref{eqn:Dc}) is ambiguous.  They showed that if the
derivative is defined via adiabatically shifting the state which is the ground
state at zero field, then the Drude weight results.  In the presence of level
crossings the adiabatically shifted state may be an excited state for a finite
value of the perturbation.  The superfluid weight is obtained if the
derivative corresponds to the ``envelope function'', i.e. the ground state of
the perturbed sytem is taken to define the derivative.  The distinction
between these two derivatives can be implemented by embedding the periodic
system under study in a larger periodic system, and defining the perturbation
in terms of the periodic boundary conditions of this larger system.  In cases
in which level crossings are close to $\Phi=0$ conductors, superconductors,
and insulators can be distinguished~\cite{Scalapino92,Scalapino93,Hetenyi12}.
In general, the position of level crossings depends on
dimensionality~\cite{Scalapino92,Scalapino93}.

The finite temperature extension of $D_c$ has been given by Zotos, Castella,
and Prelov\v{s}ek\cite{Zotos96} (ZCP).  This generalization can be summarized
as
\begin{equation}
D_c(T) = \frac{\pi}{L}\sum_n \frac{\exp(-\beta E_n)}{Q} \left[ \frac{\partial^2 E_n(\Phi)}{\partial
    \Phi^2} \right]_{\Phi=0}.
\label{eqn:ZCP_Dc}
\end{equation}
Note in this expression the Boltzmann weight factors remain unchanged as the
perturbation $\Phi$ is turned on.  Eq. (\ref{eqn:ZCP_Dc}) has been
applied~\cite{Kirchner99} to calculate the DC conductivity in strongly
correlated systems.  Taking the zero temperature limit reproduces Kohn's
expression for $D_c$.  To define a finite temperature analog of $D_s$ one lets
the Boltzmann weight factors depend on the perturbing field $\Phi$ as
\begin{equation}
D_s(T) = \frac{\pi}{L} \left[\frac{ \partial^2}{\partial \Phi^2} \sum_n
\frac{\exp(-\beta E_n(\Phi))}{Q} E_n(\Phi) \right]_{\Phi=0}.
\label{eqn:ZCP_Ds}
\end{equation}
Indeed the ground state superfluid weight is reproduced in the zero
temperature limit.  Eqs. (\ref{eqn:ZCP_Dc}) and (\ref{eqn:ZCP_Ds}) follow from
the assumption that the distinction between the Drude and superfluid weights
is due to the different types of derivatives as discussed by SWZ.

Similarly, in deriving expressions for $D_c$ and $D_s$ in a variational
setting our starting assumption will also be that the distinction between the
two quantities is due to the effects of level crossings.  Suppose
$|\tilde{\Psi}(\gamma)\rangle$ is a variational wavefunction, where $\gamma$
denotes a set of variational parameters, which we wish to use to optimize some
Hamiltonian $\hat{H}$ with eigenbasis
\begin{equation}
\hat{H}|\Psi_n \rangle = E_n |\Psi_n \rangle.
\end{equation}
The estimate for the ground state energy may be written in terms of a density
matrix as
\begin{equation}
\langle \tilde{\Psi}(\gamma)|\hat{H}|\tilde{\Psi}(\gamma)\rangle = \sum_n
\langle \tilde{\Psi}(\gamma)|\Psi_n\rangle E_n \langle \Psi_n
|\tilde{\Psi}(\gamma)\rangle = \sum_n P_n E_n,
\end{equation}
the probabilities can be written as
\begin{equation}
P_n = |\langle \tilde{\Psi}(\gamma) | \Psi_n \rangle|^2.
\end{equation}
Comparing with Eq. (\ref{eqn:ZCP_Dc}) it is obvious that a consistent
formalism requires that the variational Drude weight be defined as
\begin{equation}
D_c = \frac{\pi}{L}\sum_n P_n \frac{\partial^2 E_n(\Phi)}{\partial \Phi^2},
\end{equation}
with $P_n$ independent of the perturbation $\Phi$.  It follows that the
variational parameter $\gamma$ is independent of the perturbation $\Phi$.  The
variational analog of $D_s$ (based on Eq. (\ref{eqn:ZCP_Ds})) corresponds to
\begin{equation}
D_s = \frac{\pi}{L}\frac{\partial^2 }{\partial \Phi^2} \sum_n P_n(\Phi) E_n(\Phi),
\label{eqn:Ds}
\end{equation}
where the probabilities $P_n(\Phi)$ depend on $\Phi$ and the variational
parameters $\gamma$ in this case {\it also depend} on $\Phi$.

A central difficulty in calculating $D_c$ in a variational theory is the fact
that it depends on the exact eigenvalues of the perturbed Hamiltonian (see
Eq. (\ref{eqn:Dc})), however variational theories are usually applied in cases
where the exact solution is not easily accessible.  While the Drude weight
remains a difficult problem in general, it is shown below that the current can
be cast in terms of a geometric phase, and evaluated even in a variational
context.

\section{Current in terms of a geometric phase}

\label{sec:J}

In this section we consider the adiabatic current response of a system in
general, not only in a variational context.  After showing that the persistent
current can be expressed as a geometric phase~\cite{Berry84,Shapere89}, we
explicitly construct the mathematical tools to calculate it, and use the
results in the next section to interpret the GWF.  Since the current can be
cast in terms of observables, it follows that the calculation of the Drude
weight is also accessible, being the first derivative of the current as a
function of the Peierls phase.

Consider a system periodic in $L$, and experiencing a perturbation in the form
of a Peierls phase $\Phi$.  Its Hamiltonian can be written as
\begin{equation}
\hat{H}(\Phi) = \sum_{i=1}^N\frac{(\hat{p}_i+\Phi)^2}{2m} + V(x_1,...,x_N).
\end{equation}
The following identity also holds
\begin{equation}
\label{eqn:pH}
\partial_\Phi \hat{H}(\Phi) = \sum_{i=1}^N \frac{(\hat{p}_i+\Phi)}{m}.
\end{equation}
The ground state energy can be written as
\begin{equation}
E(\Phi) = \langle \Psi(\Phi)|\hat{H}(\Phi)| \Psi(\Phi)\rangle.
\end{equation}
The average current for such a system can be expressed as~\cite{Kohn64}
\begin{equation}
J(\Phi) = \partial_\Phi E(\Phi) = 
\langle \Psi(\Phi)|\partial_\Phi \hat{H}(\Phi)|\Psi(\Phi)\rangle.
\end{equation}
Substituting for the partial derivative of the Hamiltonian we obtain
\begin{equation}
J(\Phi) = \frac{N \Phi}{m}+ \sum_{i=1}^N
\langle \Psi(\Phi)| \frac{\hat{p}_i}{m}|\Psi(\Phi)\rangle.
\end{equation}
In the position representation the current can be written
\begin{equation}
J(\Phi) = \frac{N \Phi}{m} - \frac{i}{m} \sum_{i=1}^N
\langle \Psi(\Phi)| \frac{\partial }{\partial x_i}|\Psi(\Phi)\rangle.
\end{equation}
Next we rewrite the wavefunction in terms of a shift of the total position and
define a wavefunction
\begin{equation}
\langle x_1,...x_N| \Psi(\Phi;X) \rangle = \Psi(x_1+X,...,x_N+X;\Phi).
\end{equation}
The action of the total momentum can then be cast in terms of the derivative
with respect to $X$ as
\begin{equation}
\sum_{i=1}^N \frac{\partial}{\partial x_i}
\Psi(x_1+X,...,x_N+X;\Phi)
= \partial_X \Psi(x_1+X,...,x_N+X;\Phi).
\end{equation}
The effect of $X$ on the particle positions is similar to the effect of the
Peierls phase on the momenta.  Like the Peierls phase it is an external
parameter, so one can perform adiabatic cycles as a function of it.  Averaging
in $X$ over the unit cell $\frac{1}{L} \int_0^L \mbox{d} X...$ leads to
\begin{equation}
J(\Phi) = \frac{N\Phi}{m}-\frac{i}{mL} \int_0^L \mbox{d}X 
\langle \Psi(\Phi;X)| \partial_X| \Psi(\Phi;X)\rangle.
\label{eqn:Jphi}
\end{equation}
The second term in Eq. (\ref{eqn:Jphi}) is a geometric phase~\cite{Berry84}.
Since it results from the periodicity of the parameter $X$ it is similar to
the geometric phase derived by Zak~\cite{Zak89}.  It is also similar to the
geometric phase expression appearing in the modern theory of
polarization~\cite{Resta94}, with the variable $X$ playing the role of the
crystal momentum in this case.  Thus the current due to a perturbation can be
expressed in terms of a constant proportional to the number of particles, and
a geometric phase term.  Below an interpretation of the phase term is given.
It is interesting to note that a finite persistent current is in principle
possible for an unperturbed system (the case $\Phi=0$).

The next question to address is the actual calculation of this quantity.  We
can construct a scheme which is in the spirit of the total position operator
proposed by Resta~\cite{Resta98,Resta99} to calculate the polarization.  We
consider the case $\Phi=0$ (and suppress the notation), without loss of
generality.  We first rewrite the Berry phase appearing in the expression for
the current in terms of its discretized analog as~\cite{Resta96}
\begin{equation}
\label{eqn:JBP}
J(0) = \lim_{\Delta X \rightarrow 0} \frac{1}{mL} \mbox{Im} \ln \prod_{s=0}^{M-1}
\langle \Psi(s\Delta X) |\Psi((s+1)\Delta X) \rangle.
\end{equation}
The continuous expression can be recovered by Taylor expanding the
wavefunction $|\Psi((s+1)\Delta X)\rangle$ around $s\Delta X$ and taking the
limit as $\Delta X \rightarrow 0$.  Indeed
\begin{equation}
J(0) = \lim_{\Delta X \rightarrow 0} \frac{1}{mL} \mbox{Im} \sum_{s=0}^{M-1}
\ln \langle \Psi(s\Delta X) |\left[ |\Psi(s)\Delta X) \rangle + \partial_X
  |\Psi(s\Delta X) \rangle \Delta X \right] = 
\ln [ 1 + \langle \Psi(s\Delta X) | \partial_X | \Psi(s\Delta X) \rangle \Delta X].
\end{equation}
When the limit $\Delta X \rightarrow 0$ is taken the natural logarithm can be
expanded and we obtain
\begin{equation}
J(0) = \frac{1}{mL} \mbox{Im} \left[ \int \mbox{d}X
  \langle\Psi(X)|\partial_X|\Psi(X)\rangle \right]
 = -\frac{i}{mL} \int \mbox{d}X \langle\Psi(X)|\partial_X|\Psi(X)\rangle .
\end{equation}
The shift in the total position of the system by a value of $\Delta X$ can be
accomplished using the total position shift operator $\hat{U}(\Delta X)$.  The
explicit form of this operator will be derived below, for now we assume its
existence.  We define it as
\begin{equation}
\hat{U}(\Delta X) |\Psi(X) \rangle = |\Psi(X + \Delta X) \rangle,
\label{eqn:U}
\end{equation}
Using Eq. (\ref{eqn:U}) we can express the product in Eq. (\ref{eqn:JBP}) as
\begin{equation}
\prod_{s=0}^{M-1} \langle \Psi(s\Delta X) |\Psi((s+1)\Delta X) \rangle
= \langle \Psi(0) |\hat{U}(\Delta X)|\Psi(0) \rangle^M
\end{equation}
Substituting into Eq. (\ref{eqn:JBP}) the expression for the current becomes
\begin{equation}
\label{eqn:JBP_DX}
J(0) = \lim_{\Delta X \rightarrow 0} \frac{1}{m}\frac{1}{\Delta X}
\mbox{Im} \ln \langle \Psi(0) |\hat{U}(\Delta X)|\Psi(0) \rangle.
\end{equation}

The total position shift operator can be constructed using real space
permutation operators.  This derivation has been given
elsewhere~\cite{Essler05}, here we emphasize the main results.  In second
quantized notation the permutation operator between two positions can be
written as
\begin{equation}
P_{ij} = 1 - (c_i^\dagger - c_j^\dagger)(c_i - c_j).
\end{equation}
This operator has the properties
\begin{equation}
P_{ij}c_j = c_i P_{ij}, P_{ij}c_i = c_j P_{ij},
P_{ij}c_j^\dagger = c_i^\dagger P_{ij}, P_{ij}c_i^\dagger = c_j^\dagger P_{ij}.
\end{equation}
Assuming a grid with spacing $\Delta X$, using $P_{ij}$ we can construct an
operator which shifts all the positions on the grid in a periodic system.  The
operator
\begin{equation}
\hat{U}(\Delta X) = P_{12} P_{23}...P_{L-1L},
\end{equation}
where it is assumed that the indices refer to particular grid points, has the
property that
\begin{equation}
  \hat{U}(\Delta X)  c_i = \left\{ \begin{array}{rl}                                                                                                        
  c_{i-1}\hat{U}(\Delta X), & i = 2,...,L
  \\                                                                                                              
  c_{L}\hat{U}(\Delta X),   & i = 1.                                                                                                                      
       \end{array} \right.                                                                                                                       
\label{eqn:UU}
\end{equation}                                                                                                                                               
It also holds that
\begin{equation}
  \hat{U}(\Delta X) \tilde{c}_k = e^{i \Delta X k} \tilde{c}_k
  \hat{U}(\Delta X),                                                                             
\label{eqn:Uc}                                                                                                                                                 \end{equation}
where $\tilde{c}_k$ denotes the annihilation operator in reciprocal space.
Eq. (\ref{eqn:Uc}) can be demonstrated by Fourier transforming $\tilde{c}_k$
and applying (\ref{eqn:UU}).  Taking the Fermi sea 
\begin{equation}
|FS \rangle = \tilde{c}_{k_1}^\dagger ... \tilde{c}_{k_N}^\dagger |0\rangle,
\end{equation}
as an example one can show that
\begin{equation}
\hat{U}(\Delta X)|FS \rangle = e^{i \Delta X K}|FS \rangle,
\end{equation}
with $K = \sum_{i=1}^N k_i$.  

As an example we consider again the non-interacting Fermi sea given by
\begin{equation}
|FS\rangle = \tilde{c}_{k_1}^\dagger...\tilde{c}_{k_N}^\dagger|0\rangle,
\end{equation}
where the $k$-vectors are spread symmetrically around zero.  Applying a
perturbation $\Phi$ shifts all $k$-vectors by $\Phi$.  The resulting current
is 
\begin{equation}
J(\Phi) = \frac{2N}{m}\Phi,
\end{equation}
corresponding to a Drude weight of $D_c = 2N/m$.  It is interesting to see
that the current is proportional to {\it twice} the number of particles.  In a
Fermi sea conduction can occur due to particles as well as holes, of which at
half-filling there are an equal number.  For systems with bound particles and
holes, $J(\Phi)$ is reduced, as bound excitons do not participate in
conduction and reduce the effective number of charge carriers.  Thus the
geometric phase in Eq. (\ref{eqn:Jphi}) accounts for exciton binding.  When
all the particles are bound to holes then the constant term in
Eq. (\ref{eqn:Jphi}) is cancelled by the phase term leading to $J(\Phi)=0$.
An example of bound particles and holes in the same band leading to insulating
behaviour is the Baeriswyl variational wavefunction~\cite{Baeriswyl00}.

\section{Contribution of the geometric phase to the current response of projected wavefunctions}

In this section we provide the response theory of some commonly used projected
wavefunctions~\cite{Gutzwiller63,Gutzwiller65,Baeriswyl86,Baeriswyl00}.  We
emphasize that it is the contribution of the phase term to the current
response we calculate, not the Drude weight, which is the first derivative of
the current response with respect to the perturbing phase.

The Gutzwiller wavefunction~\cite{Gutzwiller63,Gutzwiller65} (GWF) was
proposed as a variational wavefunction for the Hubbard model, and it has the
form
\begin{equation}
|\Psi_G(\gamma)\rangle  = e^{-\gamma \hat{D}}|FS\rangle,
\end{equation}
where $\hat{D} = \sum_i n_{i\uparrow}n_{i\downarrow}$.  Without loss of
generality we consider the one-dimensional case.

Before developing the current response theory of the GWF, we present the
calculation of a quantity which expresses the extent of localization.
Localization has been suggested long ago as a general criterion of
metallicity~\cite{Kohn64}, and the relation of the spread to the DC
conductivity has been shown in a number of
places~\cite{Resta98,Resta99,Hetenyi12}.  In particular we calculate the
normalized spread defined as
\begin{equation}
\frac{\langle X^2 \rangle - \langle X \rangle^2}{L^2}.
\label{eqn:X2} 
\end{equation}
Due to the ill-defined nature of the position operator in periodic systems we
choose a sawtooth representation which can be written as
\begin{equation}
  X = \sum_{\stackrel{m=-L/2}{m\neq0}}^{L/2-1} \left(\frac{1}{2} +
  \frac{\hat{W}^m}{\mbox{exp}\left( i\frac{2\pi m}{L}\right)-1} \right),
\label{eqn:X_st} 
\end{equation}
where $\hat{W}$ denotes the total momentum shift operator, which has the
property that
\begin{equation}
\hat{W} |\Psi(K)\rangle = |\Psi(K+ (2\pi)/L)\rangle.
\end{equation}
The construction~\cite{Hetenyi09} of this operator is analogous to the total
position shift operator used to define the persistent current in section
\ref{sec:J}.  For a state $|\tilde{\Phi}\rangle$ diagonal in the position
representation one can write
\begin{equation}
\hat{W}|\tilde{\Phi}\rangle = e^{i\frac{2\pi}{L}\sum_i \hat{x}_i}|\tilde{\Phi}\rangle
\end{equation}
where $x_i$ denotes the position of particle $i$.  Using the sawtooth
representation one can show that for the Fermi sea that the spread in position
\begin{equation}
  \frac{\langle X^2 \rangle - \langle X \rangle^2}{L^2} = \lim_{L\rightarrow \infty}
\frac{1}{L^2}\sum_{m=1}^{L-1} \frac{1}{2(1-\mbox{cos}\left(\frac{2\pi
    m}{L}\right))} 
= \frac{1}{12}. 
\label{eqn:sst} 
\end{equation}
To show this one needs to substitute Eq. (\ref{eqn:X_st}) into
Eq. (\ref{eqn:X2}), and then use the fact that
\begin{equation}
  \hat{W}^m|FS\rangle = \left\{ \begin{array}{rl}                                                                                                        
  0 & m = 2,...,L-1
  \\                                                                                                              
  |FS\rangle   & m = 0.                                                                                                                      
       \end{array} \right.                                                                                                                       
\end{equation}     

Our results are shown in Table \ref{tab:spread_gwf}.  The GWF results for two
different values of the variational parameter were calculated for a
one-dimensional system based on the variational Monte Carlo method of Yokoyama
and Shiba~\cite{Yokoyama87}.  The fact that the normalized spread approaches a
constant for large $L$ (system size) indicates that the system is delocalized,
hence metallic.  What is surprising in these results, however, is that for
large $L$ the spread of all three examples converges to the same value.  The
projecting out of double occupations in the GWF seems to have no effect on the
spread for large $L$, and is identical to the result for the Fermi sea.  The
GWF though is thought to be a representative of ``bad metals'', metals whose
conductivity is reduced due to strong
correlations.~\cite{Brinkman70,Fazekas99}
\begin{table}
\begin{tabular}{|c||c|c|c|}
\hline
$L$ & Fermi sea & $\gamma=1.0$& $\gamma=2.0$  \\ \hline
$12$ & $0.08275$ & $0.079(1)$  & $0.0412(9)$   \\
$24$ & $0.08312$ & $0.0830(6)$ & $0.0682(6)$     \\
$36$ & $0.08327$ & $0.0831(5)$ & $0.0797(5)$     \\
$48$ & $0.08330$ & $0.0833(4)$ & $0.0824(4)$     \\
$60$ & $0.08331$ & $0.0829(3)$  & $0.0830(3)$     \\
$\infty$ & $1/12$ &   &  \\
\hline
\end{tabular}
\caption{Spread in the total position divided by the square of the system size
  for the Fermi sea and the Gutzwiller wavefunction.  Two different values of
  the \ variational parameter, $\gamma=1.0$ and $\gamma=2.0$ are shown.  As
  the system size increases the value $1/12$ is approached by all three
  systems.  The approach to the limiting value slows down as correlation
  effects are introduced, it is slowest for $\gamma=2.0$, the "most projected"
  of the three examples.}
\label{tab:spread_gwf}
\end{table}           

It turns out that these results are actually consistent with what one obtains
for the current response.  We consider the phase term under a perturbation in
the form in Eq. (\ref{eqn:JBP_DX}).  Consider first the action of the operator
$\hat{U}(\Delta X)$ on the GWF.
\begin{equation}
\hat{U}(\Delta X)|\Psi(\gamma)\rangle  = \hat{U}(\Delta X)e^{-\gamma \sum_i n_{i\uparrow}n_{i\downarrow}}|FS\rangle.
\end{equation}
The operator $\hat{U}(\Delta X)$ shifts the positions of {\it all particles}
by one lattice spacing.  Such a shift will not change the total number of
double occupations, hence the Gutzwiller projector and the total position
shift operator commute.  We can write
\begin{equation}
\hat{U}(\Delta X)|\Psi(\gamma)\rangle = e^{-\gamma \sum_i
  n_{i\uparrow}n_{i\downarrow}}\hat{U}(\Delta X)|FS\rangle = e^{i\Delta X
  \sum_i k_i}|\Psi(\gamma)\rangle,
\end{equation}
where $\sum_i k_i$ denote the sum over the momenta of the {\it Fermi sea}.
One obtains exactly the same result in the absence of the Gutzwiller
projector.  When substituting back into Eq. (\ref{eqn:JBP_DX}) we find that
the current response of the GWF is exactly that of the Fermi sea, and this
result is independent of correlations (whose strength increases monotonically
with the variable $\gamma$).  The above derivation can be extended to
projections based on Jastrow-type correlations and the conclusion is valid as
long as the projections are centro-symmetric (considered in
Ref. \cite{Millis91}).  It has been shown~\cite{Capello05} that
non-centrosymmetric correlators can produce an insulating state.  The current
response in this case will also not follow the derivation above, since a shift
in all the particles can change the contribution to the projector.  Another
exception is the case when $\gamma\rightarrow\infty$, i.e. the singular case,
which in general is also known to allow for insulating
behavior~\cite{Capello06}.

For the GWF one can obtain further insight into the current response by
writing it in the position representation as
\begin{equation}
|\Psi_G \rangle = \sum_{\bf R} e^{-\gamma D({\bf R})} \mbox{Det}({\bf
  K};{\bf R}) |{\bf R} \rangle.
\label{eqn:Psi_G_r}
\end{equation}
In Eq. (\ref{eqn:Psi_G_r}) ${\bf R}$ indicates the configurations of particles
(both up-spin and down-spin), $D({\bf R})$ indicates the number of double
occupations for a particular configuration of particles, $\mbox{Det}({\bf
  K};{\bf R})$ denotes the product of Slater determinants for up-spin and
down-spin electrons, and $|{\bf R} \rangle$ denotes a position space
eigenstate.  From Eq. (\ref{eqn:Psi_G_r}), we see that the projection changes
the relative weight of different configurations but leaves their phases {\it
  intact.}~\cite{Fazekas99} The fact that the current, a quantity related to
the phase of the wavefunction, is unaltered by the Gutzwiller projection
coincides with the result above, namely that the persistent current for a Gutzwiller
wavefunction is determined exclusively by the Fermi sea.  In fact Millis and
Coppersmith~\cite{Millis91} suggest a scheme in which a projector operator of the form
$e^{iS}$, with $S=\frac{1}{U}(H_t^+-H_t^-)$, ($H_t^+$($H_t^-$) raises(lowers)
the number of double occupations) acts on the Fermi sea to produce an
insulating wavefunction.  Clearly this scheme would alter the phases of the
Fermi sea.

The above reasoning can be extended to other commonly used projected
variational wavefunctions.  The Baeriswyl-Gutzwiller wavefunction can be
written
\begin{equation}
|\Psi_{BG}(\alpha,\gamma)\rangle = e^{-\alpha \hat{T}}e^{-\gamma \hat{D}}
|FS\rangle.
\label{eqn:Psi_BG}
\end{equation}
In this case $\hat{T}$ denotes the kinetic energy, and $\alpha$ denotes the
variational parameter.  Since the total position shift operator
$\hat{U}(\Delta X)$ is diagonal in momentum space, it trivially commutes with
the projector $e^{-\alpha \hat{T}}$.  We conclude that the current response of
the Baeriswyl-Gutzwiller projected wavefunction is identical to that of the
Fermi sea.  The other two commonly used variational wavefunctions are the
Baeriswyl and Gutzwiller-Baeriswyl projected wavefunctions.  Their form is
\begin{equation}
\label{eqn:Psi_B}
|\Psi_{B}(\alpha,\gamma)\rangle = e^{-\alpha \hat{T}}
|\Psi_{\infty}\rangle,
\end{equation}
\begin{equation}
\label{eqn:Psi_GB}
|\Psi_{GB}(\alpha,\gamma)\rangle = e^{-\gamma \hat{D}} e^{-\alpha \hat{T}}
|\Psi_{\infty}\rangle.
\end{equation}
In Eqs. (\ref{eqn:Psi_B}) and (\ref{eqn:Psi_GB}) $|\Psi_{\infty}\rangle$
denotes the wavefunction in the limit of infinite interaction.  This function
is in general not known.  Again one can exploit the fact that the total
position shift commutes with the projector operators and conclude that the
current response in both cases will depend on $|\Psi_{\infty}\rangle$
exclusively.  While this function is not known, in general, in the
half-filled case one can assume that its current response is zero.

\section{Conclusion}

The current response was investigated in the context of variational theory.
The Drude and superfluid weights have seemingly identical expressions (second
derivative of the ground state energy with respect to the Peierls phase),
however, as was pointed by Scalapino, White, and Zhang, the meaning of the
derivative differs between the two, one being the adiabiatic the other the
``envelope'' derivative.  Assuming their interpretation of the derivative we
derived the expressions for the Drude and superfluid weights appropriate for
variational theory.  A key difficulty with the former is the appearance of the
exact eigenstates of the perturbed Hamiltonian, in general not available in
practical situations where variational theory is used.  As a partial remedy
the persistent current was shown to consist of a constant term, proportional
to the perturbation and the number of charge carriers, and a geometric phase
term.  This expression can be used in practical settings to obtain the Drude
weight by numerically taking the first derivative of the current with respect
to the phase.  The current response of several commonly used variational
wavefunctions was also analyzed, and shown that variational wavefunctions
which use a Baeriswyl or Gutzwiller projection will have a current response
determined by the wavefunction on which the projectors are applied (Fermi sea
or the solution in the strongly interacting limit).

\section*{Acknowledgements}  This work was supported by the Turkish agency for
basic research (T\"UBITAK, grant no. 112T176).  Part of the work was carried
out at the Graz University of Technology under a grant from FWF (no.
P21240-N16).  The author is grateful to the Physical Society of Japan for
financial support in publication.

\end{document}